\documentclass[9pt,twocolumn,twoside]{pnas-new}
\usepackage{bm}
\usepackage{bbm}
\usepackage{caption}
\usepackage{subcaption}
\templatetype{pnasresearcharticle}

\title{Heterogenous criticality in high frequency finance: \\
a phase transition in flash crashes
}

\author[a,1]{Jeremy D. Turiel}
\author[a,b]{Tomaso Aste} 

\affil[a]{Department of Computer Science, UCL, Gower Street, WC1E6BT London, UK}
\affil[b]{Systemic Risk Centre, London School of Economics and Political Sciences, London, United Kingdom}
\affil[1]{Corresponding author. E-mail: jeremy.turiel@gmail.com}

\leadauthor{Turiel}

\authorcontributions{Please provide details of author contributions here.}
\authordeclaration{Please declare any conflict of interest here.}
\equalauthors{\textsuperscript{1}A.O.(Author One) and A.T. (Author Two) contributed equally to this work (remove if not applicable).}

\begin{abstract}
Flash crashes in financial markets have become increasingly important attracting attention from financial regulators, market makers as well as from the media and the broader audience. Systemic risk and propagation of shocks in financial markets is also a topic of great relevance that attracted increasing attention in recent years. In the present work we bridge the gap between these two topics with an in-depth investigation of the systemic risk structure of co-crashes in high frequency trading. 
We find that large co-crashes are systemic in their nature and differ from small crashes. We demonstrate that there is a phase transition between co-crashes of small and large sizes, where the former involves mostly illiquid stocks while large and liquid stocks are the most represented and central in the latter. 
This suggests that systemic effects and shock propagation might be triggered by simultaneous withdrawals or movement of liquidity by HFTs, arbitrageurs and market makers with cross-asset exposures.

\end{abstract}

\dates{\today}

\begin{document}

\maketitle

\ifthenelse{\boolean{shortarticle}}{\ifthenelse{\boolean{singlecolumn}}{\abscontentformatted}{\abscontent}}{}

\section{\label{intro}Introduction}

Flash crashes in financial markets can be defined as extreme changes in the price of one or multiple assets within a short interval of time. These have become increasingly relevant for practitioners and in particular market makers whilst being increasingly studied and reported in the quantitative finance literature.

%The introduction is structured in two parts. The former provides the reader with general context on flash crashes and related dynamics with a focus on HFT market players, while the latter leads on to the remainder of the paper with a more focused overview of flash crashes, systemic risk and propagation of liquidity shocks in HFT markets. Flash crashes and systemic risk are the two main themes which this work brings together. Our in depth analysis of systemic risk in flash crashes, based on real data, constitutes the work's main contribution to the literature.

The most notorious flash crash is likely that of May 6th 2010, which involved the major US stock indices (S\&P, DJIA, NASDAQ composite) and caused a $\approx 9 \%$ drop in the DJIA in the 36 minutes it lasted for. This event led to a variety of empirical and theoretical papers trying to understand the event and its causes with the aim to shed light on other black swan events too (up/down crashes).
High Frequency Traders are at the center of interest in a large portion of this literature, hence we report a brief summary of their role in markets and its regulatory concerns.

It has been shown that HFT market players contribute to price efficiency and tighter spreads, thereby improving the price discovery process. These players and electronic trading as a whole have become increasingly dominant in recent years to the point of constituting a large portion of the traded volume in financial markets.
On the other hand, some characteristics of HFT players caused other market players to raise concerns as the run to incredibly fast execution leaves many behind and allows HFTs to front run other players \cite{chlistalla2011high}. The ability of HFTs to process information faster than other players leads to adverse selection and its fixed cost to a size advantage for larger players which might hurt the overall welfare of market participants \cite{biais2011equilibrium}. It can now perhaps be argued that the run to faster execution is going beyond price efficiency which benefits investors and towards an unstable price process driven by competition between large firms.
This is supported by a large body of literature on flash crashes which places HFTs at the center of some disruptive systemic events, as discussed below.

The SEC's report on the flash crash of May 6th \cite{cftc2010findings} finds that most market participants automatically halted their trading due to hard risk constraints triggered by the sudden price change, while some HFT firms kept trading, as it was deemed still profitable by their algorithms. These absorbed most of the original large sell order, but once they reached inventory or loss constraints they started selling too. This increased the selling pressure in the market and some works hold that it caused HFTs to trade with each other repeatedly (``hot potato phenomenon'') thereby increasing the traded volume (but not the real liquidity). This apparent increase in liquidity in the form of high trading volume caused large sell orders to get executed faster \cite{golub2012high}. 
This chain of events highlights that the phenomenon has a dangerous positive feedback loop.

Results in \cite{paddrik2012agent} show through simulations how reducing either (or both) the number of HFT players or the size of the large sell order greatly reduced the size of the drawdown.
Further, other works find that black swan phenomena of duration $< 1.5s$ are about ten times more frequent than longer ones and their return distribution deviates from the canonical power law distribution of returns. The authors suggest a phase transition to an all-machine environment at $\sim 1s$ as human reaction time is in the order of seconds. The authors also investigate the time scales via additional simulations to show the rise in extreme events and their magnitude around $\sim 1s$ in what they define as the all-machine phase \cite{johnson2012financial, nanex2010nanex}.
Findings along those lines, on the distribution of High Frequency black swan events deviating from the canonical return distributions, were also recently publishes by the authors of this work \cite{turiel2021self}.

%emergence of black swan phenomena to arise as a phase transition at lower trading frequencies .

From the review above we see that crashes of different sizes seem to involve a self-perpetuating cycle \cite{paddrik2012agent} with positive feedback loops.

This type of self-excited process is also investigated in \cite{cespa2014illiquidity} for the liquidity and information dependence between two sample assets, showing how liquidity shocks to an asset can propagate to related ones (and by extension to the wider market).

The frequency and size (in terms of number of securities involved) of simultaneous-like crashes in HFT is also investigated in the literature.
For instance, the works by Lillo and co-authors \cite{calcagnile2018collective, bormetti2015modelling} investigate the dynamics of simultaneous flash crashes, and motivate their importance by showing the growth in the number of mini crashes in recent years. Further, they show how the number of simultaneously crashing securities grew over the last 10 years, thereby highlighting the increasing systemic relevance of this phenomenon.

%As we move to discuss shock propagation and systemic risk we clarify our understanding of 
In this paper we define systemic risk as the risk component of an event (say a flash crash) that is given by the interconnectedness of assets, likely as a results of correlated actions and arbitrage between market participants. This causes isolated events to spread in the market and affect more assets, thereby increasing their impact and relevance for all market participants. A related concept is that of ``synchronisation'' which is the systemic and concentration aspect that arises from the alignment and interdependence of actions between market players (on a single asset) rather than across assets.

The systemic risk posed by HFTs has been investigated in the literature in the last decade. 
The work by Paulin et al. \cite{paulin2019understanding} simulates flash crashes through agent based modeling and highlights the importance of market structures in the systemic propagation of extreme events.
The works by Abreu \& Brunnermeier \cite{abreu2002synchronization} and Bhojraj et al. \cite{bhojraj2009margin} investigate the risks of synchronisation of arbitrageurs in financial markets and acknowledges the phenomenon. Other works investigate the systemic risk of HFT dynamics. % of interest in this work,
Jain et al. \cite{jain2016does} investigate how low-latency HFT trading can worsen extreme systemic events in financial markets and argues for the need to incorporate correlation and market structure in regulating these risks. The work by Harris \cite{harris2013high} discusses many mechanisms, among which, systemic risks originating from order routing and self-reinforcing mechanisms causing crashes. The review by De Gruyter \cite{serrano2020high} summarises systemic aspects of HFTs and market structure such as position correlation and herd behaviour, adverse selection in orders and crowding as well as negative contribution to price discovery, at times.

%It was shown in the literature how, as electronic trading increasingly takes over the majority of the traded volume in markets, the number of flash crashes and the size of co-crashes has increased in recent years \cite{calcagnile2018collective}.
%The works by Lillo and co-authors \cite{calcagnile2018collective, bormetti2015modelling} investigate the dynamics of frequency and size of co-crashes as a single model, but as we see these events 
Co-crashes are becoming more frequent and systemic. 
It is therefore important to investigate their structure. % of co-crashes grows. 
In particular, it is relevant to understand which stocks are central to larger systemic events as well as the contagion structure between stocks in the market. This is a central theme in market stability for regulators as well as in risk management for market makers.

%As suggested in the introduction, 
The present work joins the two themes of flash crashes and systemic risk by delving deeper into the dynamics of simultaneous flash crashes of different sizes throughout 300 liquid stocks traded on the NASDAQ. We investigate the empirical distribution of crash sizes and the structure of these events in the market. We also investigate whether larger systemic events involve highly unstable stocks (which crash often) or stocks that are more stable in their price dynamics, yet more influential to trigger larger systemic events when subject to liquidity shocks.
We apply tools from statistical physics to show the difference between crashes which involve a small or large number of assets.
%show through null models of crash frequency how crashes of smaller sizes belong to a different dynamic than that of larger systemic events and discuss implications for systemic risk in high frequency markets.
We uncover a phase transition occurring when the crash size exceeds 5 companies.
Implications for systemic risk in high frequency markets are discussed from both trading and regulatory perspectives.

\section{\label{data}Data}

In the present work we consider a universe of 300 liquid stocks from the NASDAQ exchange between 3/1/2017 and 25/9/2020. High frequency price data is obtained from LOBSTER \cite{huang2011} and sampled to obtain non-overlapping one minute returns. This frequency was also adopted in \cite{calcagnile2018collective} and other works in the literature for the detection of price jumps as it is understood that below this limit microstructural noise becomes relevant and can impact the validity of the method.

%\section{\label{intro}Background}
\section{\label{method}Method}

\subsection{Jump Detection}

We detect price jumps (up and down crashes) similarly to \cite{calcagnile2018collective}, at least in principle, in 1 minute non-overlapping returns.

Specifically, we apply the basic jump detection method from \cite{lee2008jumps} and detect jumps at the $5\%$ significance level. In addition to the basic features of the method for robust volatility estimation we obtain a robust estimate of intraweek periodicity and adjust the return series and jump detection according to \cite{boudt2011robust}.

Null models are calibrated and price jumps detected individually for each stock. As we consider 1 minute non-overlapping returns, our sampling allows for aligned timestamps. We then consider contemporaneous price jump detection across assets in the universe as simultaneous jumps (a single systemic event).

\subsection{Crash size distribution and firm persistence}
\label{method:crash_distribution}

We investigate whether co-jumps which involve different numbers of stocks originate from the same dynamic process and present the same distribution. We also consider whether individual stocks are involved to the same extent across co-crashes of different sizes or if a pattern emerges.

%To do this we construct the following null model of co-crash components.

We define the unnormalised crash frequency for stock $x$, in co-crashes with $m$ stocks as

$$f_{x,m} = \sum_{t=0}^T c_{x,t,m}$$

with 
$$c_{x,t,m} =
\begin{cases}
      1, & \text{if stock $x$ is involved in a crash of size $m$ at time $t$} \\
      0, & \text{otherwise}
\end{cases}$$

By marginalising over the ensemble of stocks $x$ we obtain the frequency distribution across co-crash sizes 

$$
f_{m} = \sum_x f_{x,m}
$$

%For completeness the normalised probability distribution over co-crash sizes can be defined as

%$$p_m = \frac{f_{m}}{\sum_{m} f_{m}} = \frac{\sum_x f_{x,m}}{\sum_{x,m} f_{x,m}}$$

The changes in the composition of the crashes are investigated by computing the correlation between the involvement of firms across crashes of different sizes.
Namely, for each crash size $m$ we assign to each firm $x$ a rank in decreasing order by $f_{x,m}$. 
We then compute the Spearman correlations between these ranks. 

\subsection{Statistical testing}

To support the visual intuition of our results we apply statistical testing in the form of null models. We have applied the Spearman correlation to test for rank similarity between the crash frequency distributions across stocks at different crash sizes $m$. As the frequency distributions are noisy and fat-tailed the more correlation p-value seems hard to justify as a valid test.
Hence, we follow the idea of Mantegna et al. \cite{curme2015emergence} to create a simple null model of correlation significance.

To do so we sample without replacement the whole list of stocks $S_m$ according to $\propto f_m$ from Section \ref{method:crash_distribution} to obtain a biased reshuffling $G_{i,m}$ of the stocks according to their crash frequency.

For each shuffled list we calculate the Spearman correlation coefficient between the sample and the original list to form the null distribution as:

$$\operatorname{D}_{m} = {Spearman(G_{i,m}, S_m)}_{i=1}^{10^5}$$

We then define the significance of the correlation between sizes $m, m+ \tau$ as the quantile of $Spearman(S_{m + \tau}, S_m)$ in $\operatorname{D}_{m}$.

\subsection{Crash-weighted trading volume}\label{sec:method:volume_crash_corr}

To investigate the relationship between crash size and the involvement of highly traded stocks we define a weighted average daily Dollar Traded Volume for each crash size, where the weighting is given by the normalised crash frequency of each stock.

For crash size $m$ and crash frequency distribution $f_{x,m}$, as per Section \ref{method:crash_distribution}, we define the crash-weighted Dollar Traded Volume $\operatorname{DTV}_m$ as:

$$\operatorname{DTV}_m = \frac{ \sum_{x} f_{x,m} \operatorname{DTV}_{x, m} } { f_m } $$

This measure aims to represent how more highly traded stocks are involved at different crash sizes.

\section{\label{results}Results and Discussion}

\begin{figure}[h]
     \centering
     \begin{subfigure}[b]{1\linewidth}
         \centering
         \includegraphics[width=\linewidth]{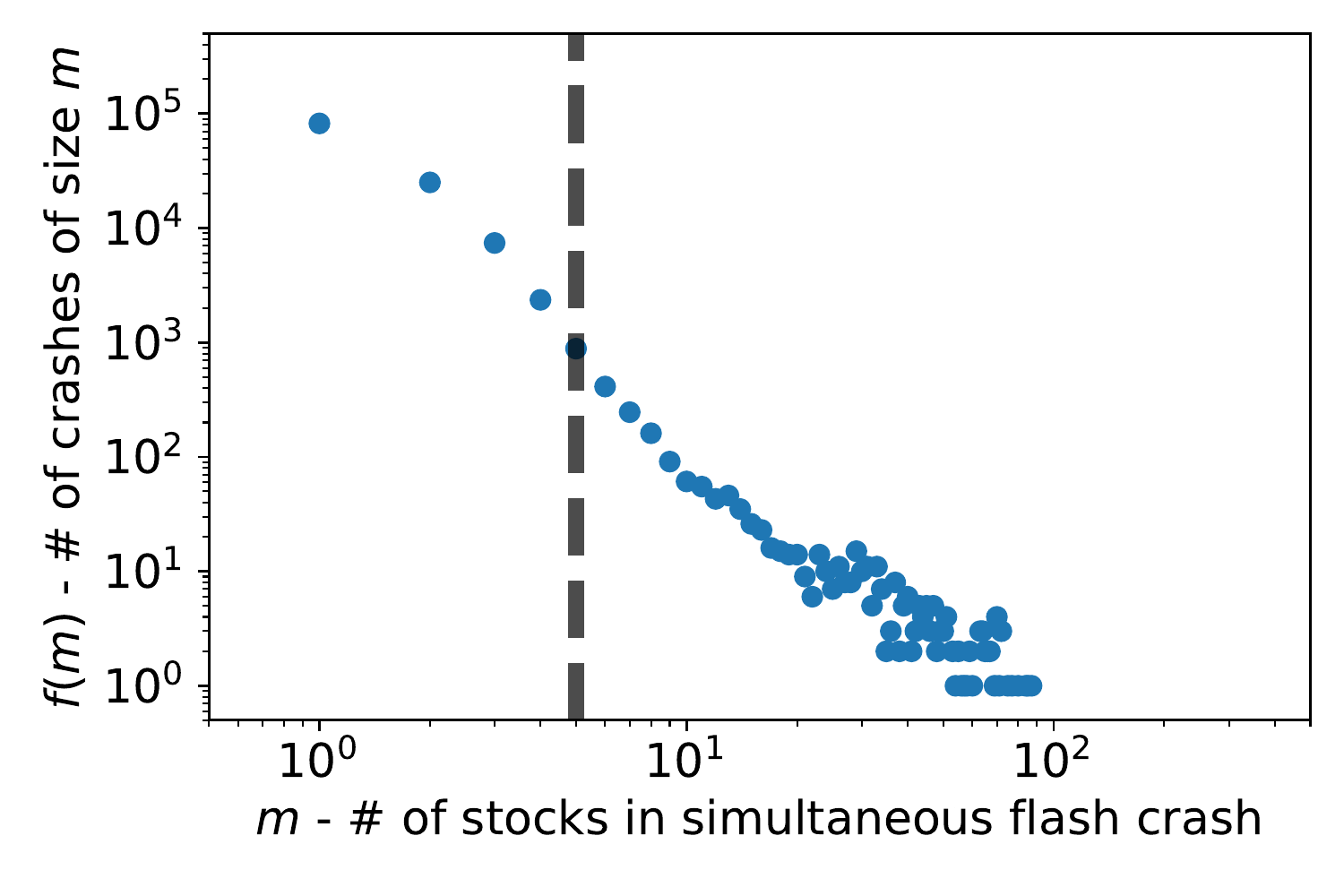}
         \caption{$f(m)$}
         \label{fig:crash_size_dist}
     \end{subfigure}
     \hfill
     \begin{subfigure}[b]{1\linewidth}
         \centering
         \includegraphics[width=\linewidth]{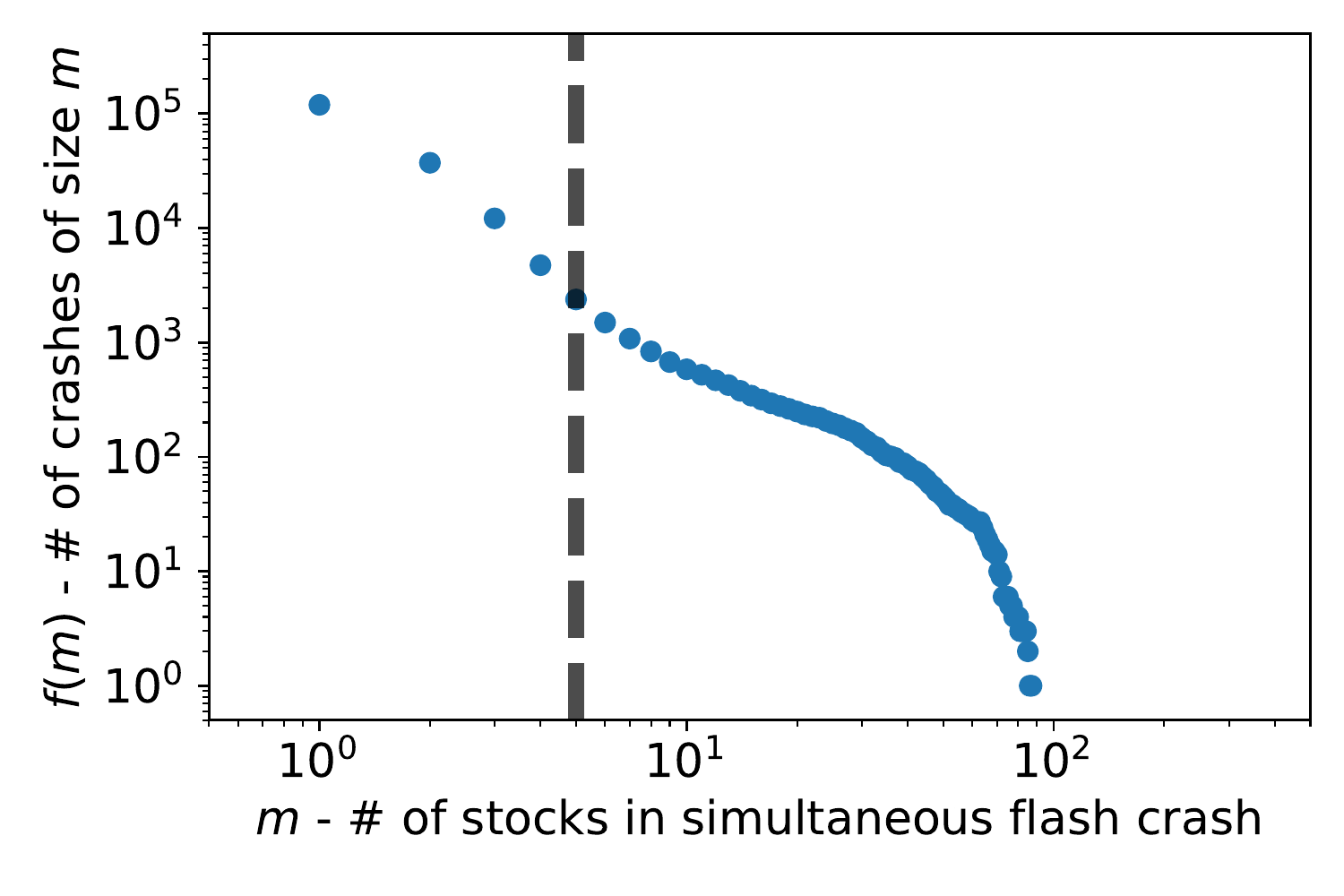}
         \caption{$f(M \geq m)$}
         \label{fig:crash_size_dist_cdf}
     \end{subfigure}
        \caption{{\bf Heterogeneous crash distribution} Log-log plot of the flash crash size distribution. We observe that sizes lesser than 4 follow a different trend, with lower than expected frequency. This suggests that crashes of this size and onwards do not belong to the same self-organised process, but that this is rather a heterogeneous distribution.}
        \label{fig:crash_size_distributions}
\end{figure}

The plot in Figure \ref{fig:crash_size_dist} shows the frequency distribution $f_m$ of the number of stocks involved in each flash crash. Figure \ref{fig:crash_size_dist_cdf} plots the cumulative frequency $f(M \geq m)$. 
It is evident from both figures that they are  heavy-tailed and there is a change in the slope around $m\approx 5$ and a finite size effect at $\approx 10^2$ which is when the crash involves a large portion of the system (system size $3\cdot 10^2$) \cite{christensen2005complexity}. This kind of distribution was already reported in \cite{calcagnile2018collective}, where the authors investigated and modelled flash crash sizes and frequency as a single Hawkes process. 
The authors there suggest that each security's crash dynamics should be modeled as a self-excitation process, but they point out that this would involve tuning a large number of parameters on very noisy data. They therefore decide to model the collective self-excitation process of securities as the frequency of crashes (or co-crashes) and their size. Hence, all crash sizes are treated as instances of a multi-asset Hawkes process in \cite{calcagnile2018collective}, with no distinction between the assets involved in each crash or their structure.

In the present work we take a more granular approach and move to investigate the structure of co-crashes and the individual susceptibility of each stock.

\begin{figure}
     \centering
     \begin{subfigure}[b]{1\linewidth}
         \centering
         \includegraphics[width=\linewidth]{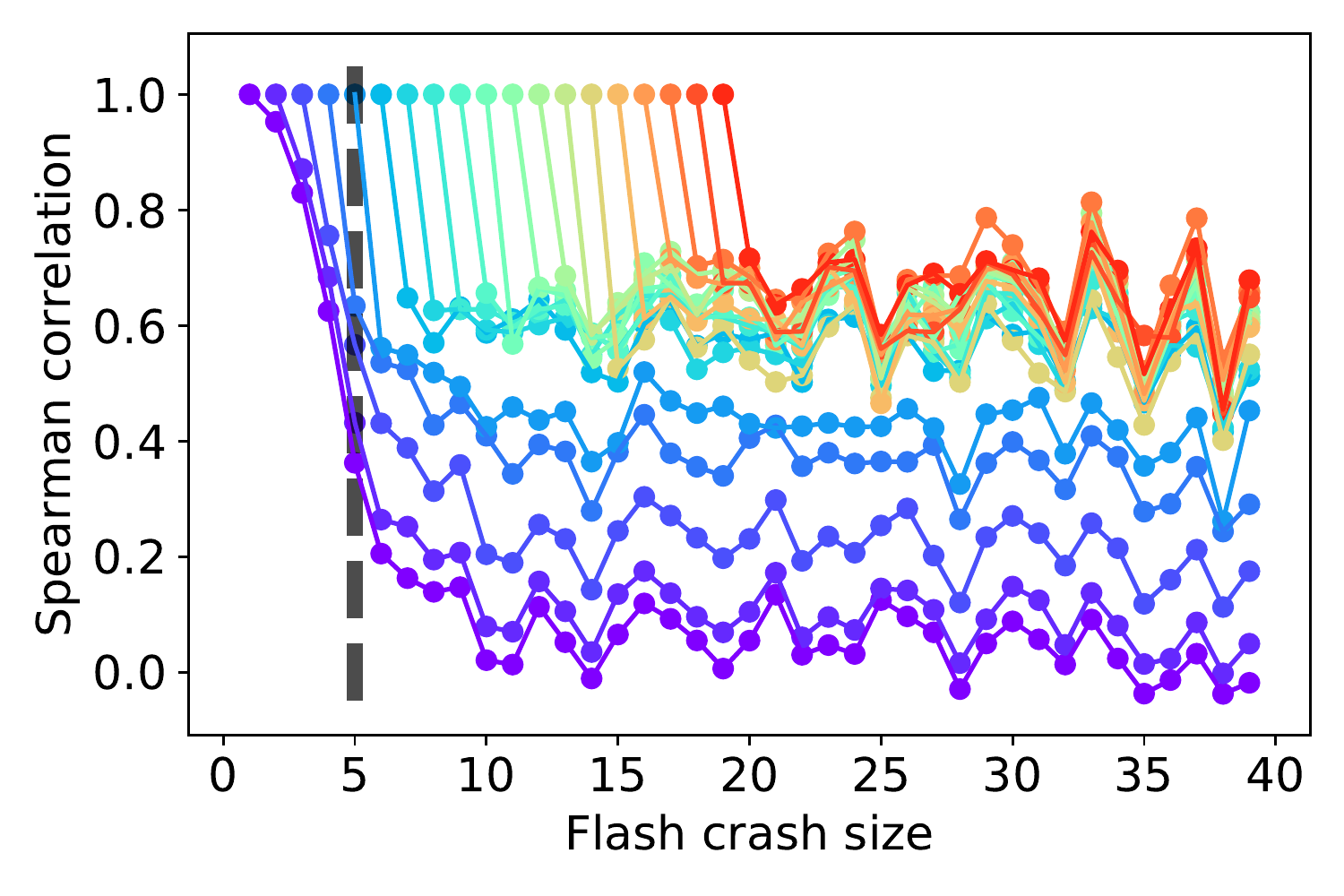}
         \caption{Spearman correlation between all consecutive crash sizes.}
         \label{fig:rank_crash_corr_raw}
     \end{subfigure}
     \hfill
     \begin{subfigure}[b]{1\linewidth}
         \centering
         \includegraphics[width=\linewidth]{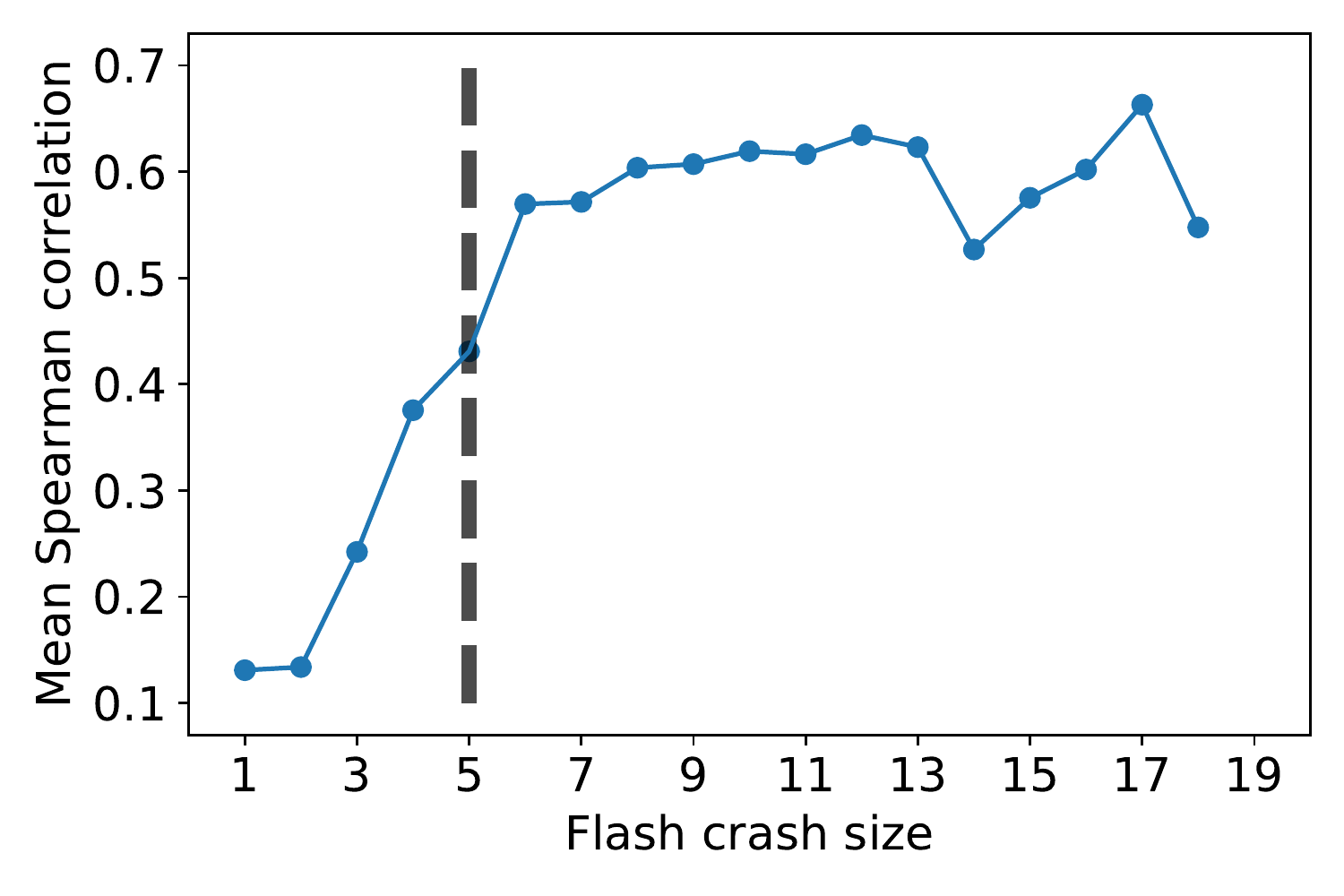}
         \caption{Spearman steady state correlation mean in $[m+2, m+20] \, , \, \forall m$.}
         \label{fig:rank_crash_corr_steady_state}
     \end{subfigure}
        \caption{{\bf Crash component rank correlation}
    Evidence that there is a transition around $m=5$ with crashes involving a small number of companies ($m<5$) being substantially differently populated with respect to crashes involving a larger number of companies ($m>5$). 
    The plot in Figure \textbf{a} reports the Spearman correlations of ranks in frequency between each starting crash size and higher crash sizes. The plot in Figure \textbf{b} looks at the average correlation in the range $[m+2, m+20]$ for each value of $m$ from Figure $a$, which offers better visual intuition.}
        \label{fig:rank_crash_corr}
\end{figure}

To further investigate the difference between small and large crash sizes we report in Figure \ref{fig:rank_crash_corr_raw} the Spearman correlation between the ranks of % $Kendall_\rho ( \{f_{x,m}\}_{x \in stocks}, \{f_{x,m+\tau}\}_{x \in stocks} )$ of 
crash frequency for all stocks.
Specifically, each line reports the correlation between the between rank of the companies in the initial crash of size $m$ (correlation 1) with all other crashes with higher sizes $m+\tau$. We indeed observe how crashes of smaller sizes ($m<5$) have a substantially different composition to crashes of larger sizes. 
%do not transition immediately to the steady state, but rather decay slowly to reach the steady state at size 6. 
We instead observe that for sizes above 6 a steady state is reached with a large component of the population represented across all crashes with similar ranks in frequency. 
%Further, steady states with starting points up to size 6 are distinct from higher ones which decay immediately to very close steady state levels. 
These steady states for $m>5$ are significantly higher than the ones of smaller sizes, as the structure no longer evolves significantly between higher size crashes. The plot in Figure \ref{fig:rank_crash_corr_steady_state} provides a clearer visualisation of this. We highlight that already at size 5 the correlation transitions directly to the steady state, although a lower one with respect to the ones for crash size 6 and above. 

To validate the visual results from Figure \ref{fig:rank_crash_corr} we apply the null model of correlation significance between crash frequency distributions.
%Figure \ref{fig:null_model_transition}  the full distribution as per Figure \ref{fig:rank_crash_corr} yields a more nuanced, albeit consistent, view of the phase transition. 
%Figure \ref{fig:null_model_transition} shows the occurrence of a very clear transition between $m=4$ and $5$ with a statistically significant steady presence of top-ranked stocks for crashes larger than $m=5$.
%how already crashes of size $\approx3-4$ are significantly more indicative than smaller ones.

\begin{figure}
\includegraphics[width=1\linewidth]{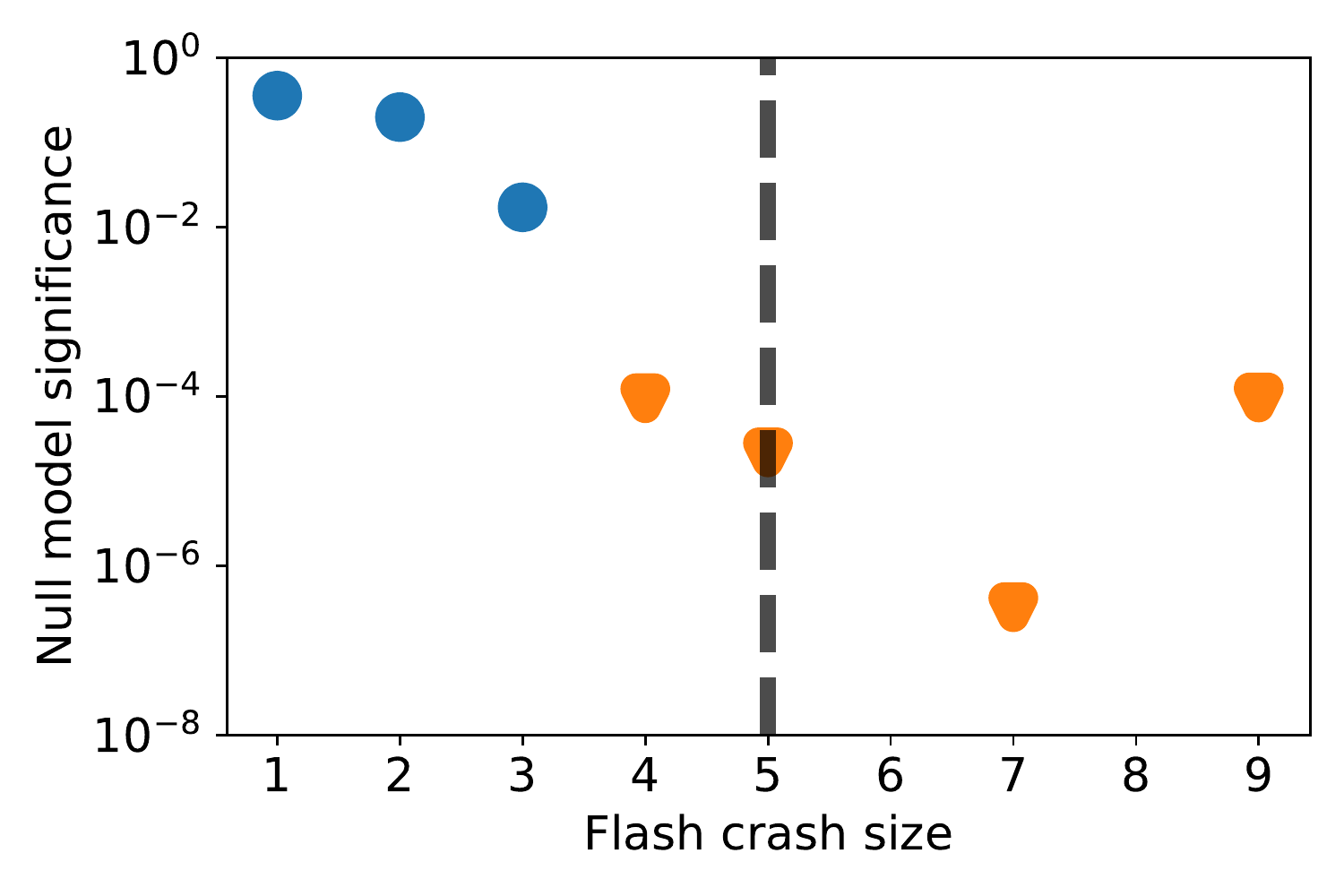}
\caption{{\bf Crash component significance phase transition}
Evidence of a transition in the dynamics of crashes composition occurring around $m=5$. 
The plot reports the steady state statistical significance of the base crash size's frequency distribution.}
\label{fig:null_model_transition}
\end{figure}

Figure \ref{fig:null_model_transition} shows the correlation significance between the starting point $m$ on the horizontal axis and its steady state distribution $\sim [m+2,m+10]$. We observe the first significant value at $1\%$ around $m=4$ which confirms the intuition from Figures \ref{fig:crash_size_dist}, \ref{fig:crash_size_dist_cdf} that crash sizes up to $\approx 4$ belong to a different process than larger crashes. Indeed smaller crashes are dominated by less stable stocks and larger ones by very liquid stocks with high market capitalisation. This suggests that more influential and systemic stocks are involved in larger crashes and perhaps even trigger those. A reason for why this is not the case in small crashes can be that these stocks are systemic enough to mostly be involved in (or perhaps even cause) crashes of larger size. These are then even more relevant for systemic risk.
%Results from a more complex null model support these conclusions and are reported in Appendix \ref{appendix:crash_null_model} for sake of simplicity.

This is therefore further evidence of the occurrence of a transition in the process between smaller and larger crashes. The slow decay of smaller crash sizes indicates how these belong to similar distributions of non-systemic events, but as the crash size grows the steady state gets closer to the large crash level. This suggests that larger crashes have some systemic characteristics.

If we take a closer look at the top ranked stocks at each size we observe that smaller crash sizes are dominated by very volatile and illiquid stocks which are subject to large jumps perhaps due to the lack of a smooth price process in their trading. We would expect this though to make them susceptible to larger systemic events as well and hence stably ranked. Yet, we observe very low to null rank correlation between individual (and small) crash frequencies and the large crash size steady state. It seems as if not only these crashes are non-indicative, but also as indicated by the phase transition in Figure \ref{fig:null_model_transition} they belong to an unrelated ranking and distribution. We highlight that we considered rankings and ranking correlation in order to avoid any sensitivity to large values or outliers at smaller frequencies.

Large crash sizes involve stocks such as Microsoft (MSFT) and Apple (AAPL) as consistently high ranked. We highlight that these stocks are highly liquid and characterised by a stable price process with very few price jumps. Indeed the few times they get involved into jumps they are often part of larger simultaneous crashes, which involve more stable and systemic stocks. Further, when analysing the co-crash relations between pairs of stocks we observed a heavy-tailed distribution of centrality for these large systemic stocks which suggests a community and core-periphery like structure of the contagion network of co-crashes \cite{latora2017complex, rombach2014core, everett2005extending, barucca2016centrality, da2008centrality}.

The above observations prompted us to conduct further analyses on the relation between stock liquidity (where average daily Dollar traded volume is used as a proxy) and crash frequency at different crash sizes.

\begin{figure}
     \centering
     \begin{subfigure}[b]{1\linewidth}
         \centering
         \includegraphics[width=\linewidth]{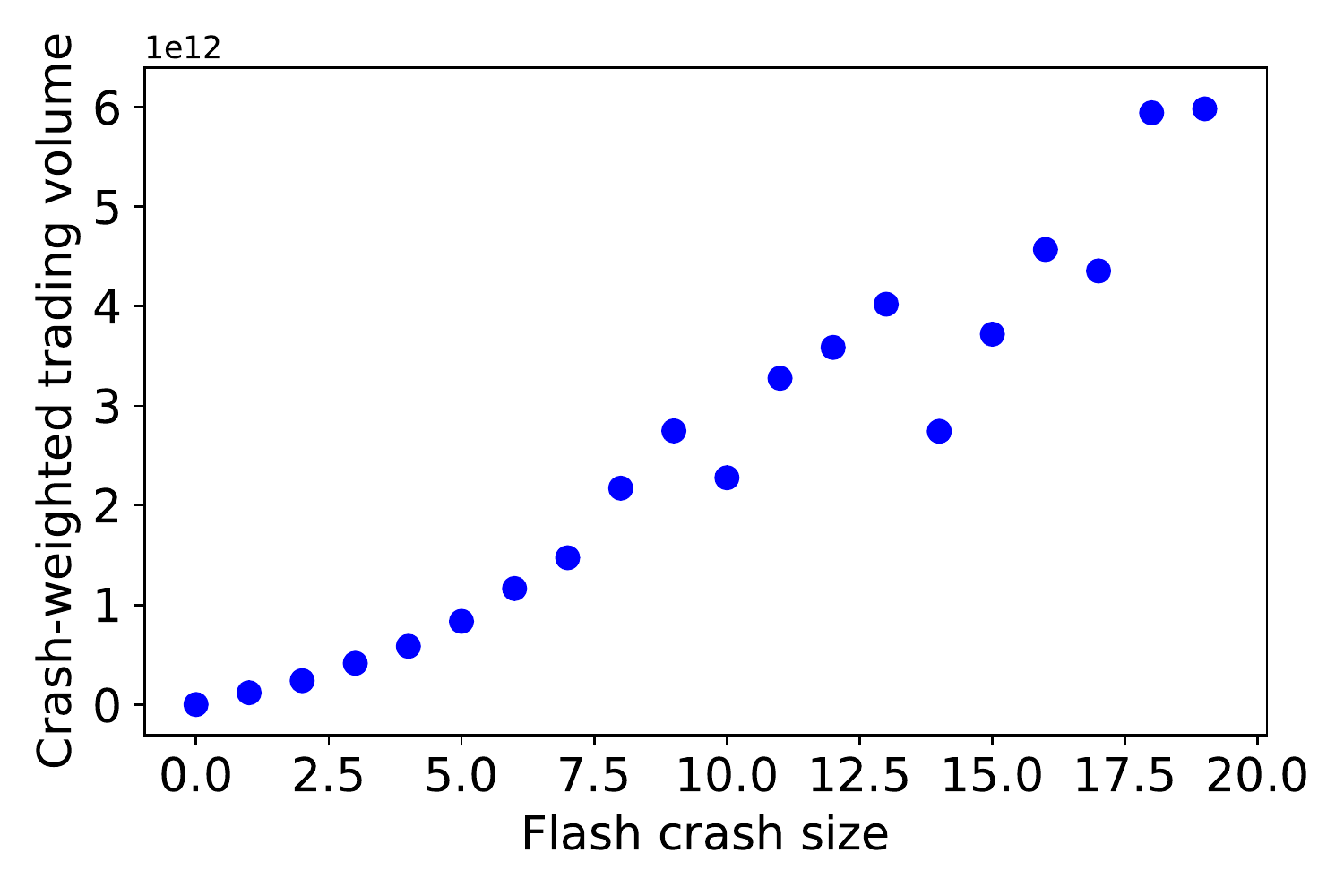}
         \caption{Positive relation between crash-weighted average daily Dollar Traded Volume and crash size $m$.}
         \label{fig:crash_weighted_vol_vs_crash_size}
     \end{subfigure}
     \hfill
     \begin{subfigure}[b]{1\linewidth}
         \centering
         \includegraphics[width=\linewidth]{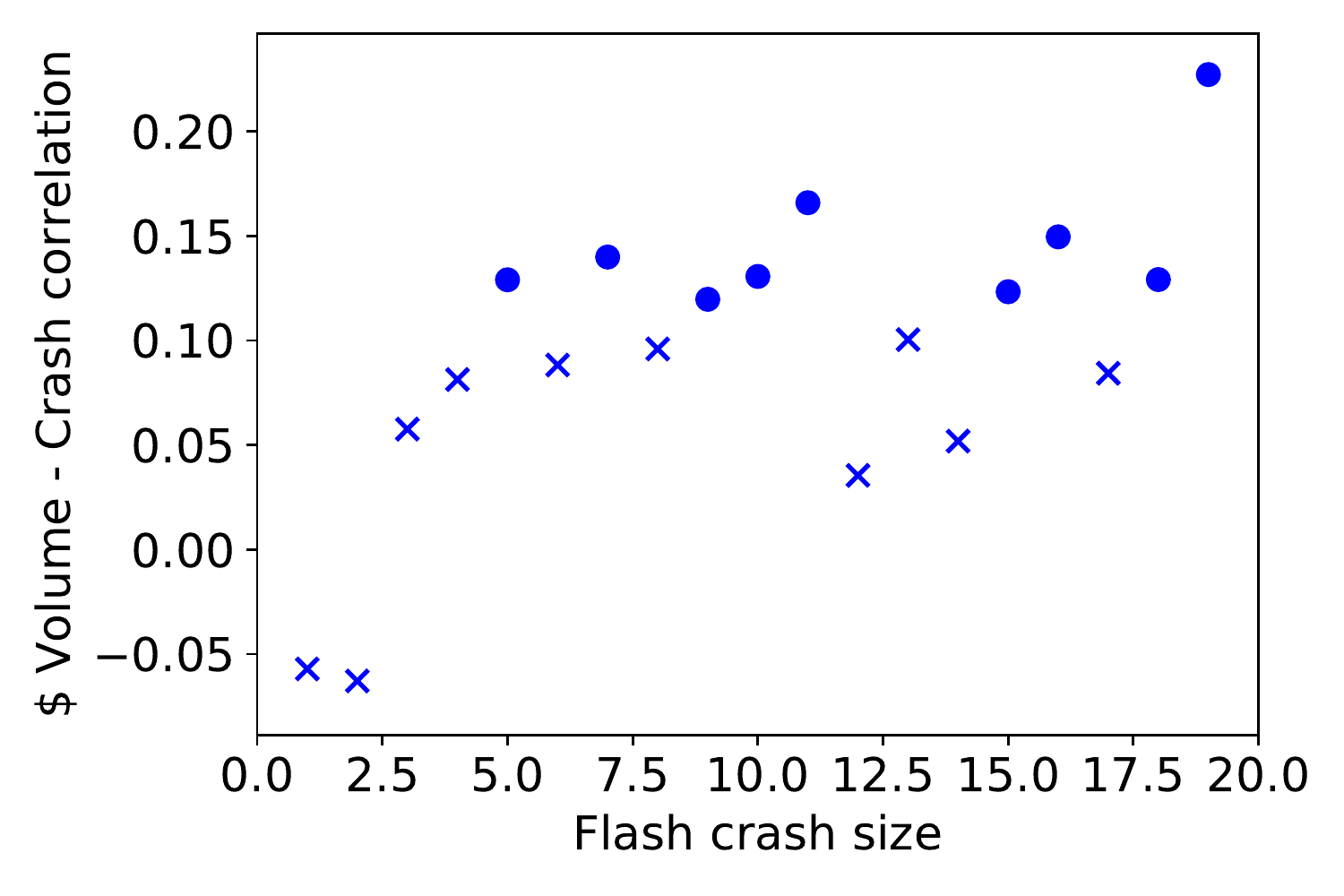}
         \caption{Spearman correlation between traded volume and crash frequency across crash sizes $m$.}
         \label{fig:volume_crash_corr_vs_crash_size}
     \end{subfigure}
     \begin{subfigure}[b]{1\linewidth}
         \centering
         \includegraphics[width=\linewidth]{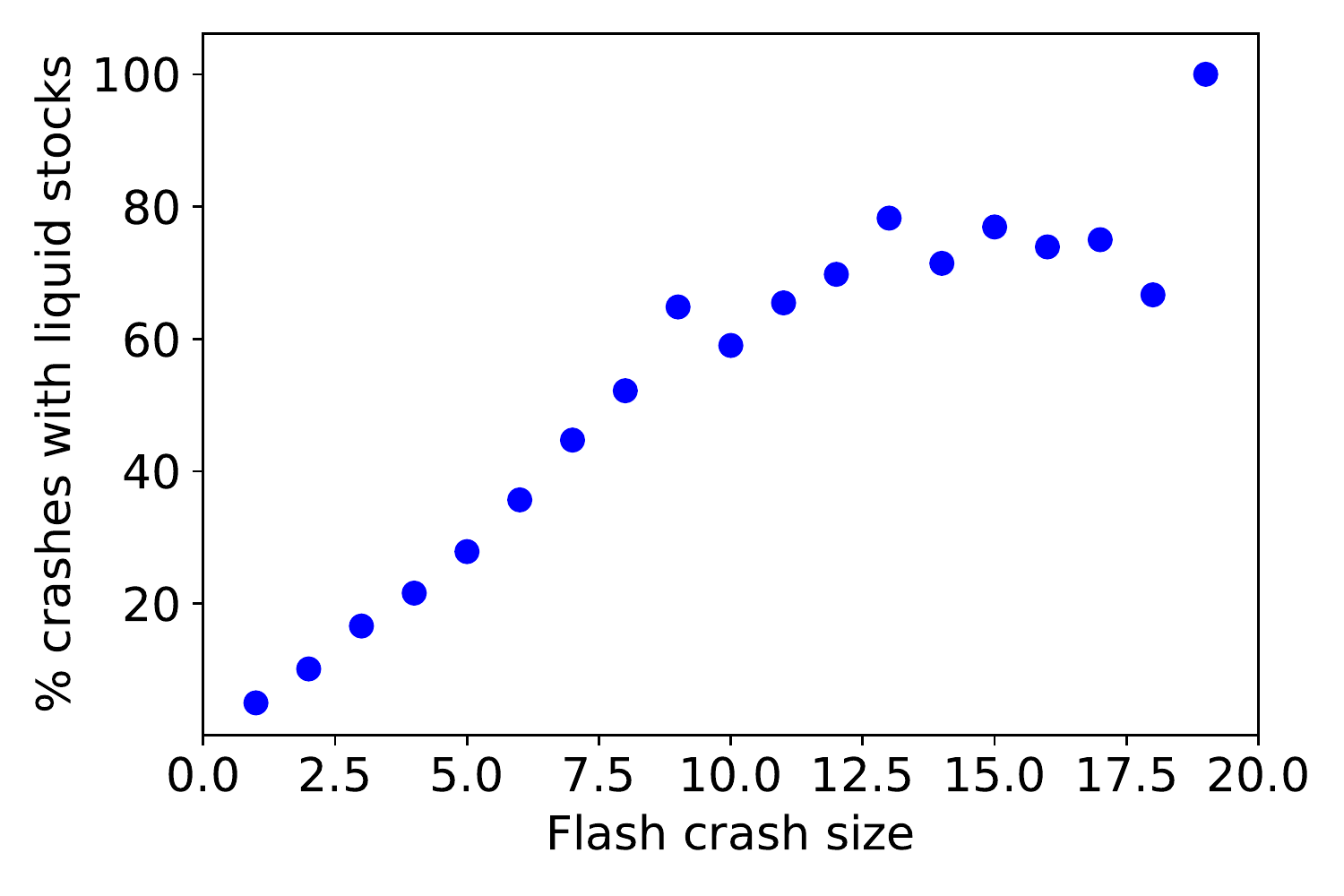}
         \caption{Positive relation between fraction of crashes involving liquid stocks and crash size $m$.}
         \label{fig:crashes_with_liquid_stock}
     \end{subfigure}
        \caption{{\bf Relation of Traded Volume to crash size.}
    The figures above show evidence of a relationship between the traded volume of stocks and their involvement in crashes of different sizes. Figure \textbf{(a)} shows the general positive relation between crash size and involvement of highly traded stocks. Figures \textbf{(b), (c)} show how the relationship exists not only on average, but also how ``liquid'' stocks are more involved throughout crashes at higher crash sizes.}
        \label{fig:volume_crash_relation}
\end{figure}

To validate visually and numerically our observation that highly traded stocks are more present in large crashes we present the plots in Figure \ref{fig:volume_crash_relation}. The plot in Figure \ref{fig:crash_weighted_vol_vs_crash_size} shows the average daily traded volume of a stock per crash size, weighted by its crash frequency, as per the definition in Section \ref{sec:method:volume_crash_corr}. This is plotted against the crash size to show a clearly increasing trend in crash-weighted traded volume with crash size. This shows how larger crashes see stocks with higher traded volumes more frequently involved.

This could though be the consequence of a subset of crashes which involved highly traded stocks. We therefore test this with results in Figure \ref{fig:crashes_with_liquid_stock} which show how not only the average crashing stock is more ``liquid'' in larger crashes, but also that the fraction of crashes which involve at least one of the top 20 stocks by traded volume in our universe increases with crash size.

In line with this, we test how the traded volume of each stock correlates with its crash frequency, for each crash size. We report results for the Spearman correlation coefficient in Figure \ref{fig:volume_crash_corr_vs_crash_size}, where dots are used for correlations significant to the $5\%$ confidence level and crosses otherwise. We see that co-crashes of size 1 and 2 seem to have an inverse or no relation between volume traded and crash size. At our previously identified phase transition point $m \approx 5$ we see the first significant positive correlation between volume traded and crash frequency which stays somewhat stable or is slightly increasing with crash size.

This last result is less clear than the previous one, but still shows a positive correlation between volume traded (a proxy for liquidity) and crash frequency at crash sizes $m > 5$.

The presence of liquid stocks in most large crashes observed in Figure \ref{fig:crashes_with_liquid_stock} prompts questions around the periphery structure of the different liquid stocks and implications for systemic risk. Further work in this direction is already underway with promising results and will be the topic of a followup work. The causality of such co-crash structures is also a very important topic, albeit harder to investigate rigorously, and should be the subject of future work.

\section{\label{conclusion}Conclusion}

The present work analyses co-jump structures in High Frequency markets. We investigate the distribution of co-jump sizes for 300 stocks on 1 minute returns. We highlight features of this distribution such as the finite size effect in the tail and the divergence of small crash frequencies from the distribution. 
We %then apply a null model of crash frequency at each crash size to 
show how the ranking and structure of crash frequency throughout stocks changes drastically through a phase transition between small and large crash sizes at size 5. We quantify this with the Spearman correlation between crash frequency ranks at different crash sizes. 
We then  apply a null model of crash frequency at each crash size to test the hypothesis of a phase transition.
Finally, we highlight how larger crashes are dominated not by the less liquid stocks present in small crashes, but rather by highly liquid stocks which are present in most flash crashes as the crash size grows. Preliminary results, which we leave for future work, find these stocks to be systemic in communities and core-periphery like structures of co-crashes. We suggest that these systemic events can be viewed as communities centered around these most influential stocks. % perhaps by means of ETFs.

We know from the literature that these structures can be indeed vulnerable and highly unstable, as well as fragmented if characterised by multiple cores. One of the possible reasons for this can be inferred from the interviews with different market players following the crash of May 6th \cite{cftc2010findings}. Many HFTs highlight the centralised risk constraints for volatility and P\&L which cause them to withdraw from the market in case of extreme conditions or losses. As they constitute much of the liquidity in the market in particular for smaller stocks, withdrawing from those causes liquidity draughts. These are often systemic as players have central risk constraints and withdraw from the entire market as those are triggered. Further, as systemic stocks crash arbitrageurs come into play to level prices across the market, hence making the isolated event a systemic one. In this view well-known stocks are not systemic per se, but rather as a result of non-siloed trading by HFTs and ETFs.

In light of the present results future works shall investigate the asyncronous price changes of securities and model spreading dynamics of flash crashes and their directed structure. Lead-lag investigations of causality of these larger crashes are also suggested for future work. Already from our results one can monitor in particular the most systemic stocks from larger flash crashes for co-jumps of size 5 and higher and induce trading halts or limitations to avoid further spreading of these systemic events. This is crucial as our results combined with those of \cite{calcagnile2018collective} suggest a systemic self-excited process in both frequency and magnitude of those crashes.

We leave the investigation of this structure for future work and highlight that this is of high importance for practitioners and regulators when dealing with market efficiency and stability, particularly as trading frequencies rise and electronic trading grows widespread across securities.

We conclude by observing that volatility and P\&L-based trading breaks used by market players may worsen these events and their systemic characteristics since they cause liquidity withdrawals throughout stocks and market players. This introduces systemic synchronisation throughout the market and makes individual assets more susceptible to small trading volumes. Further, we suggest to monitor the stocks we find systemic throughout larger crashes to model the contagion of liquidity crises and halt trading before these spread and distort a larger number of assets. This should also be topic of future work aimed at smart and efficient regulation in High Frequency markets.

\section{Acknowledgements}

JT acnowledges Riccardo Marcaccioli for useful discussions and support with the jump detection method.
TA and JT acknowledge the EC Horizon 2020 FIN-Tech project for partial support and useful opportunities for discussion. JT acknowledges support from EPSRC (EP/L015129/1). TA acknowledges support from ESRC (ES/K002309/1), EPSRC (EP/P031730/1) and EC (H2020-ICT-2018-2 825215).

\section{Author Contributions}

JT conducted the computational analysis and drafted the manuscript. TA guided the work and interpretation of results and reviewed and edited the manuscript.

\section{Conflict of Interest}

The authors declare no conflict of interest.

\section*{References}

\bibliography{main}

\end{document}